\documentstyle[prl,aps]{revtex}
\begin{document}
\draft
\tighten
\twocolumn[\hsize\textwidth\columnwidth\hsize\csname @twocolumnfalse\endcsname
\title{ Passive scalars, random flux, and chiral phase fluids }
\author{Jonathan Miller and Jane Wang}
\address{ James Franck Institute, The University of Chicago,
Chicago, IL 60637}
\maketitle

\begin{abstract}

We study the two-dimensional localization problem for
(i) a classical diffusing particle advected by a quenched
random mean-zero vorticity field, and (ii) a
quantum particle in a quenched random mean-zero magnetic field.
Through a combination of numerical and analytic
techniques we argue that both systems have extended eigenstates
at a special point in the spectrum, $E_c$, where a sublattice
decomposition obtains. In a neighborhood of this point,
the Lyapunov exponents of the transfer-matrices acquire
ratios characteristic of conformal invariance allowing an indirect
determination of $1/r$ for the typical spatial decay of
eigenstates.

\end{abstract}
\vskip2pc] \narrowtext

In this paper we study two simple models for passive advection
of a diffusing field:  (I) the diffusion of a scalar density,
$n({\bf x})$, advected by a quenched random velocity field,
${\bf A}({\bf x})$, described by the Fokker-Planck
equation\cite{bouchaud}:
$$
        {\partial_t n} = {\cal L}_{fp} n \equiv
        D \nabla^2 n - {\nabla} \cdot ({\bf A} n)
        \eqno\hbox{(I)}
$$
where $D$ is the diffusivity; and (II) the random-flux
model\cite{numerical} for a non-interacting quantum particle
propagating in a spatially-random, zero-mean magnetic field $B
\equiv\nabla \times {\bf A} \equiv \partial_x A_y -
\partial_y A_x$, where ${\bf A}$ now denotes the vector
potential, described by the Schrodinger equation
$$
        { -i\partial_t \psi} = {\cal L}_{rf} \psi \equiv
        \lbrace ({\bf p} - {\bf A})^2 + V
        \rbrace \psi
        \eqno\hbox{(II)}
$$
$\psi$ being the (complex) quantum wave function,
${\bf p} \equiv - i \nabla$ the momentum operator,
and $V({\bf x})$ the (scalar) potential disorder\cite{altshuler}.

Model (II) has received much attention recently in the
context of the quantum Hall effect at filling
factor $\nu={1 \over 2}$\cite{numerical}.
An unresolved question is whether the system has
properties of a Fermi liquid, and in particular,
extended states.  Previous work has addressed the
energy-dependence of the localization length, $\xi(E)$,
moving inward from the band edge, with authors
arriving at opposite conclusions. The most careful numerical
study of model (II) to date concludes that all states
are localized\cite{numerical}$(b)$, whereas others find a central
{\it band} of extended states\cite{numerical}$(a)$.  An analytic
calculation using a replicated nonlinear sigma model
with topological term\cite{arovas} also obtains a band
of extended states.

Our purpose is two-fold. First, we explore the
consequences of particle-hole symmetry at the band
center of these models, $E_c$, and describe
numerical and analytical evidence for a divergent localization length at
this point.  Previous studies\cite{aronov}\cite{numerical}$(b)$
did not allow for this symmetry at the band center\cite{ludwig}.
Second, we demonstrate that the properties of random flux
that have drawn so much attention are exhibited by a much larger
class of models, among them the passive scalar model (I).

Magnetic field and vorticity are distinguished from
potential fields by their transformation under time-reversal
\cite{graham}. Writing the velocity in model (I) as
${\bf A} = \nabla \chi + \nabla \times \phi$, we observe
that the curl-free part, $\nabla \chi$, like
$V$, is even under time-reversal, whereas the
divergence-free part $\nabla \times \phi$ (the source of vorticity
$\omega = \nabla \times {\bf A} = - \nabla^2 \phi$), like $B$,
is odd. A further physical similarity between
the two models is that one expects transport to be
dominated by the longest streamlines\cite{fisher,trugman};
for the random-flux (passive scalar) model with
vanishing mean magnetic field (vorticity), these rare
streamlines run along the interfaces of opposing magnetic
field (vorticity). The Laplacian enables fields to tunnel
(diffuse) among distinct closed streamlines, and creates
competition between {\it advection} by streamlines
that can transport the field coherently over long distances,
and {\it diffusion} that leads to destructive interference\cite{susy}.

We study spatial decay of the eigenfunctions
for lattice approximations to ${\cal L}_{fp}$ and ${\cal L}_{rf}$.
${\cal L}_{fp}$ is not self-adjoint, and its eigenvalues, $z$,
occupy in general an {\it area} in the complex plane.  For
given $z$ we use well-established numerical transfer-matrix
methods and finite-size scaling\cite{vulpiani} to compute
the localization length\cite{thouless} on a long strip.

The real scalar field $n$ is discretized on a square lattice
of width $L$, length $m$.  Various choices for boundary
conditions in the transverse ($L$) direction will be discussed
later. $n$ and $\chi$ are defined on nodes, ${\bf A}$ on links,
and $\phi$ on the nodes of the dual lattice.
We define lattice difference operators $\Delta_{\mu}^+ \phi_{\bf x}
\equiv \phi_{{\bf x}+{\bf e}_\mu}-\phi_{\bf x}$ and
$\Delta_{\mu}^- \phi_{\bf x} \equiv \phi_{\bf x}
-\phi_{{\bf x}-{\bf e}_\mu}$, where ${\bf e}_\mu$, $\mu=x,y$,
are orthogonal lattice basis vectors.  The
velocity is then ${\bf A} = \Delta^- \times \phi + \Delta^+ \chi$,
where $\Delta^- \times \phi \equiv (-\Delta_y^- \phi,\Delta_x^- \phi)$
represents the discrete curl.  Defining the
current ${\bf J} = D \Delta^+n - {\bf A}n$, where
$[{\bf A}n]^\mu_{\bf x} \equiv {1 \over 2}A^\mu_{\bf x}
[n_{\bf x} + n_{{\bf x}+{\bf e}_\mu}]$ represents an average of
the two ends of the link, the equation of motion is
$\partial_t n + \Delta^- \cdot {\bf J} = 0$.

Similarly, to discretize the random-flux Hamiltonian we
define the lattice covariant derivative:
$$
        \displaylines{
        D^+_{\mu} \psi_{\bf x} \equiv e^{i A^\mu_{\bf x}}
        \psi_{{\bf x}+{\bf e}_\mu} - \psi_{\bf x} \cr
        {\cal L}_{rf} \equiv D^+_\mu D^-_\mu + V_{\bf x}}
$$
The vector potential ${\bf A}$ has been defined on links
in the same way as for the fluid; the scalar potential
$V_{\bf x}$ vanishes unless otherwise stated.

Values of the field can be computed recursively using
the $2L\times 2L$ transfer-matrices, $W_k$, which yield
values of $\psi$ or $n$ in lattice column $k+1$ and $k$
given those in columns $k$ and $k-1$.  Lyapunov exponents
are extracted
as logarithms of the $2L$ eigenvalues of the matrix
$(W^{(m)\dagger} W^{(m)})^{1/2m}$ in the limit
$m \to \infty$, where $W^{(m)} \equiv \prod_{k=1}^m W_k$;
correlation lengths along the
strip then correspond to their inverses.  We define the
scaled localization length by $\xi_L(z) = 1/\lambda_L(z) L$,
where $\lambda_L(z)$ is the exponent smallest in magnitude.
A critical or extended phase occurs for those $z$ where
$\xi_L(z) \to \xi_\infty(z) \ne 0$ for large $L$.

We first describe results for model I for incompressible
${\bf A} = \Delta^- \times \phi$, taking the $\phi_{\bf x}$
to be independent random variables distributed uniformly over
an interval $[-w,w]$.  When $\phi \equiv 0$ the eigenvalues
fill (uniformly, in $d=2$) the real interval $[-8D,0]$. For nonzero
$\phi$ the density of states broadens into a
complex neighborhood of this interval.  Fig.~1(a) displays $\xi_L(z)$
using $D=1/4$, $w=1$, $L=32$ and periodic boundary conditions.
The peaks at $z=0,-8D$ arise
because the eigenfunctions, corresponding respectively to
$n({\bf x})=n_0$ and to uniform antiferromagnetic $n({\bf x})$
represent exact solutions for any width.
The structure is symmetric about the line
$\hbox{Re}(z) = E_c \equiv -4D$, a feature that originates in an
exact {\it particle-hole} symmetry of ${\cal L}_{fp}$. For $L$
{\it even}, we divide the lattice into into its two equivalent
antiferromagnetic sublattices, with $n_+$ and $n_-$ the
restriction of $n$ to the two sublattices, and
$\hat{\lambda} \equiv \bigl({n_+ \atop n_-}\bigr)$.  We can
now express ${\cal L}_{fp}$ in block form operating
on $\hat \lambda$:
$$
{\cal L}_{fp} = \left(
\begin{array}{c}
E_c I_L \hspace{0.5cm}  {\cal Q} \\
{\tilde {\cal Q}} \hspace{0.5cm} E_c I_L
\end{array}
\right),
        \eqno\hbox{(4)}
$$
where $I_L$ is the $L \times L$ unit matrix, $\tilde{\cal Q}
= T{\cal Q}^{\top}T^{-1}$,
$\top$ denotes transposition, and $T$ time reversal.  This
transformation inverts the sign of the anti-symmetric parts
of ${\cal L}_{fp}$.
The symmetry may be stated as follows:  if $\hat \lambda_z$ is an
eigenvector of ${\cal L}_{fp}$ with eigenvalue $z$, then
$\hat \lambda_{\bar z} \equiv \sigma_z \hat \lambda^* \equiv
\bigl({n^*_+ \atop -n^*_-}\bigr)$ is an eigenvector with eigenvalue
$\bar z = 2E_c - z^*$.  The same symmetry applies to the
random flux operator, where $E_c = 4$, $z$ is real, and
$\tilde {\cal Q}={\cal Q}^\dagger$.

At $z=E_c$ the two sublattices
decouple, and the eigenvectors of interest correspond to the
zero eigenvalues of ${\cal Q}$ and $\tilde {\cal Q}$. Furthermore,
in the limit $m \to \infty$, the Lyapunov exponents at $E_c$ must
occur in degenerate pairs; this degeneracy is obtained
numerically and disappears for any $z \neq E_c$.  The decoupling
is also the source of the striking depression in $\xi_L$ at
$E_c$ seen in Fig.~1 for both models.

The depression and the degeneracy occur only for even $L$.
For odd $L$, the boundary conditions mix the two sublattices.
For periodic boundary conditions,
$\xi_L(E)$ reaches (at $E \simeq -6.0$ for the random flux
model) a plateau of twice its degenerate value as one moves
in from the band edges, and maintains that value at the band
center.  For odd $L$ and {\it free} boundary conditions [as used
in some numerical studies\cite{numerical}$(a)$], $\xi_L(E_c)$
{\it diverges} as $m \to \infty$ for any finite $L$.  We obtain
both degeneracy and divergence also for the
``q''-models\cite{numerical}$(a,c)$ where the fluxes
are restricted to values $2 \pi n/q$ with $q,n$ integers.
These properties have not been identified before.

Numerical evidence alone cannot distinguish an infinite
localization length from a large but finite one. We now
offer analytic arguments in support of a divergent correlation
length at $E_c$ for the random-flux model with free
boundary conditions. Let $\hat n_k \equiv \bigl({n_k \atop
n_{k-1}}\bigr)$ represent the $2L$-component vector composed
of the $n_{\bf x}$ in columns $k$ and $k-1$.  With an
appropriate choice of gauge, we can write the
transfer-matrix in the form $W_k \equiv \bigl({\Theta_k \atop {I_L}}
{{-I_L} \atop 0}\bigr)$ where $\Theta_k$ is hermitian. Observe
that $W_k^\dagger
J W_k = J$ where $J \equiv \bigl({0 \atop {I_L}} {{-I_L}
\atop 0}\bigr)$;  the set of matrices satisfying this identity
constitute a group.  It follows that the eigenvalues of $W_k$
occur in inverse conjugate pairs, $\mu,1/\mu^*$, as do the
eigenvalues of any product of $W_k$'s.

The sublattice decomposition enables us to
reorganize the components of $\hat n_k$ in the form $\hat n_k =
\hbox{col}\lbrace n_k^+,n_{k-1}^+,n_k^-,n_{k-1}^- \rbrace$.
If $L=2l+1$ is odd, the number of components
of $n_k^\pm$ will alternate with $k$ between $l$ and $l+1$.
The transfer-matrix now takes the block form $W_k = \bigl(
{w_k \atop 0} {0 \atop \tilde w_k} \bigr)$ where $w_k = \bigl(
{\theta_k \atop I} {-I \atop 0} \bigr)$ and $\tilde w_k = \bigl(
{\theta_k^\dagger \atop I} {-I \atop 0} \bigr)$.
This new form of the transfer-matrix only connects sites on
the same sublattice and the $\theta_k$ are no longer hermitian
nor (for odd $L$) necessarily square.

Because the ensemble of random matrices weights
$\theta_k$ and $\theta_k^\dagger$ equally, we expect that
the eigenvalues of the submatrix products $(w^{(m)\dagger} w^{(m)})^{1/2m}$
and $(\tilde w^{(m)\dagger} \tilde w^{(m)})^{1/2m}$
are identical in the limit $m \to \infty$, where
$w^{(m)} \equiv \prod_{k=1}^m w_k$ and
$\tilde w^{(m)} \equiv \prod_{k=1}^m \tilde w_k$.  It follows
that the eigenvalues of the full transfer-matrix product
occur in degenerate pairs; this degeneracy is observed numerically.
Since the full transfer-matrix product is a group element, we
deduce that for {\it odd} $L$, there must be a pair of eigenvalues
with modulus unity, one from each of the two submatrix
products, yielding a divergent $\xi_L(E_c)$.

%
For {\it even} $L$, an eigenvalue of modulus unity
is not expected for finite $L$, and we instead argue that a pair
of eigenstates exists at $E_c$ in the thermodynamic
limit. To make further progress we
turn our attention from the transfer-matrix $W$ to
the operator ${\cal L}_{rf}$ itself, and exploit
its special form (4) at $E_c$: the singular value decomposition
of ${\cal Q}$. (For convenience, we translate the
$0$ of energy to $E_c$ in this discussion). For a lattice with an odd number
of sites, $N$, ${\cal Q}$ is not square so that it has
a non-trivial kernel and $0$ is a two-fold degenerate eigenvalue
of ${\cal L}_{rf}$. As a consequence of the singular value
decomposition, adding a new lattice site (reversing the
parity of $N$) can never increase
the magnitude of the smallest non-zero eigenvalue. Because
the randomness in ${\cal Q}$ can be expected to remove any
accidental exact or near degeneracy, we anticipate rather that
the magnitude of the smallest eigenvalue above $0$ (and its
particle-hole conjugate below $0$) diminishes as $N \to \infty$,
yielding a degenerate pair of $0$ eigenvalues.

If we accept that in the limit $N \to \infty$ through even values
there are indeed two independent eigenfunctions at $0$, they
must take the form $\bigl({u \atop {\pm v}}\bigr)$, $u,v$ the
left, right eigenvectors of $\cal Q$ so that
$Q^{\dagger}u = Qv = 0$. Because
$<u|v> \ne 0$ in general, we see that the arbitrary relative phase of
$u$ and $v$ implies a continuous $U(1)$ symmetry at $E_c$.
%
%
Such a continuous symmetry is known to play an important
role in some closely-related random-matrix models.
Wegner first observed the significance of the
sublattice decomposition in a class of random-matrix models
for localization in $d=2$\cite{gade}. It was later noticed that the sublattice
decomposition allows a new continuous symmetry, which contains in the
$n=0$ replica limit a factor of $U(1)$. For lattice models with spin,
this continuous symmetry leads to a divergent DOS and
localization length\cite{gade}. Our numerics indicate a
finite DOS at $E_c$, a result that in general is not
inconsistent with a divergent correlation length.  We are
pursuing an analogous replicated field theory calculation
for our model; results so far are consistent with the
existence of this symmetry\cite{jm}.

We turn now to the {\it neighborhood} of the band center.
Assuming finite-size scaling, we expect $\xi_L(z)$ to be
independent of $L$  (though not of $z$: see\cite{dhlee})
as $L \to \infty$ in the regime of extended states.  In
Fig.~2, we show $\xi_L(z)$ for several values of $z$
and $L$.  Numerical values for $\xi_{L}(z)$ are more or
less independent of $L$ when $z$ is on the lines $\hbox{Im}(z)=0$
or $\hbox{Re}(z)=E_c$.

For $E_c$ and its neighborhood we have examined the
entire Lyapunov spectrum, and find that after appropriate
scaling (see Fig.~3) the spectra for distinct $L$ collapse
onto a single curve.  In addition, for small $n/L$,
ratios of Lyapunov exponents take the form
$$
        \lambda^n_L/ \lambda^1_L = 2n-1
        \eqno\hbox{(5)}
$$
where $\lambda^j_L$ denotes the j-th largest positive
Lyapunov exponent.  This relation is remarkably
universal:  it continues to obtain for both models (I) and (II)
even when continued to complex ${\bf A}$ (this allows
an approximate interpolation between the two models),
and also when the $B_{\bf x}$ (rather than the
${\bf A}_{\bf x}^\mu$) in model II are
taken to be independent random variables.
As explained elsewhere\cite{dhlee}, the form $R_n
\equiv (n+x)/x$ for the ratio of Lyapunov exponents
suggests that the single-particle Green's function
is conformally invariant for a typical realization
of the disorder, and typically decays as $1/r^{\eta}$,
where $\eta=2x$. Evidently, for our models $\eta \simeq 1$.

So far we have discussed only divergence-free ${\bf A}$.
For model (I), the addition of a random $\chi$
leads to a real non-zero diagonal component
of the discretized model at $E_c$ [as does scalar potential
disorder $V_{\bf x}$ for model (II) -- note that $\chi$ may always
be removed by a gauge transformation].
The diagonal components of the disorder invalidate
particle-hole symmetry and the sublattice decomposition.
For small $w$, our numerics for these cases are nevertheless {\it consistent}
with extended states, and the form $x \simeq 1/2$ for the ratio
still obtains. For a {\it pure}
curl-free random velocity field (only $\chi$ nonzero),
we find that all states are localized.

We have done further numerical computations in order
to isolate the property of the operators
${\cal L}_{fp}$ and ${\cal L}_{rf}$ responsible for the
apparent band of extended states\cite{jm}.
It is found that
symmetric real random-matrix models with a sublattice
decomposition display at $E_c$ degenerate extended states
and $x \simeq 1/2$,
but away from $E_c$ all states are unambiguously
localized, as is observed for the $q=2$ model\cite{numerical}$(a)$,
and $R_n$ has no special form.
The addition of a random, uncorrelated anti-symmetric
matrix (real or complex) produces, in the neighborhood of $E_c$,
a band of states
for which the Lyapunov ratios satisfy $x \simeq 1/2$
and $\xi_L$ has no detectable dependence on $L$.
Recalling that the {\it symplectic} group constitutes
the set of transformations under
which a bilinear anti-symmetric form is invariant, we
conjecture that the usual assumption that the universality
class of these models is described only by a {\it unitary}
symmetry\cite{aronov}\cite{numerical}$(b)$ may not be justified.

We propose that {\it chiral phase fluids}, systems with large or
divergent correlation lengths originating in a random anti-symmetric
contribution to their dynamics, are a general phenomenon and share
universal features. A full understanding of these systems
must answer the question, not so far resolved numerically, of whether
the {\it only} extended states are in fact at the band center, and if so,
what sets the scale of the correlation length in the remainder
of the band.  Since our arguments are general in nature, we
expect to see similar universal behavior in other
systems, among them the kinematic dynamo\cite{zeldovich}.

\acknowledgements
We are grateful to E. Martinec, S. Abanov, L. Kadanoff,
P. Wiegmann, D.P. Arovas, and S-C. Zhang for advice and discussions.
We especially thank P. Weichman for extensively revising our manuscript.
This work was supported primarily by the MRSEC Program of the
National Science Foundation under Award Number DMR-9400379.
Extensive use was made of the CRAY C98/8128 at SDSC.

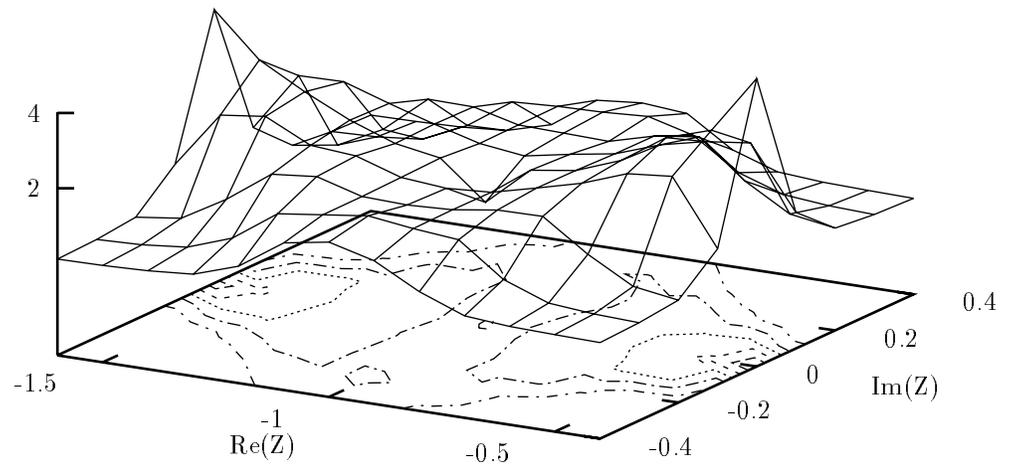
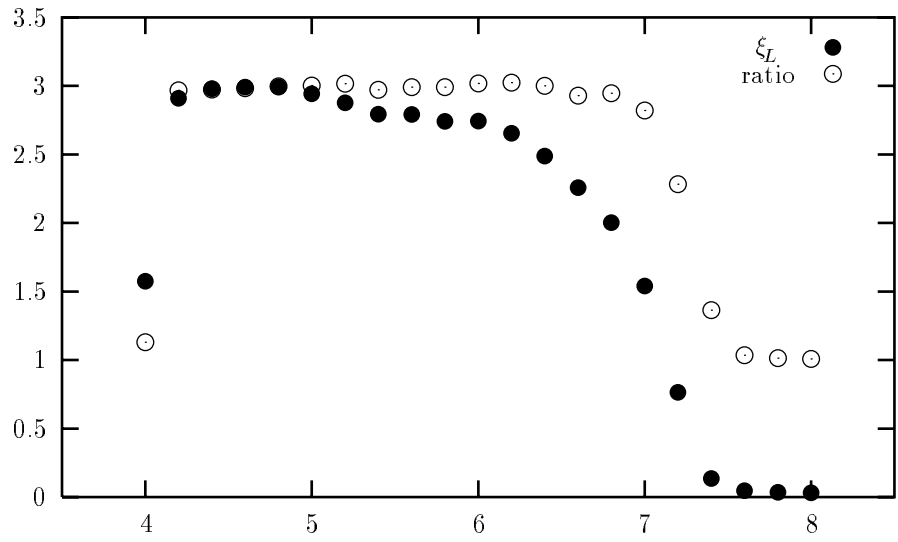
\begin{figure}
\caption{(a) Scaled localization length, $\xi_L(z)$, for model (I)
and contour plot.  Parameters are $D={1 \over 4}$, $w=1$,
$L=32$ with periodic boundary conditions.
The inset shows $\xi_L(z)$ for model (II), with $L=32$ and
fluxes chosen independently and uniformly on $[0,2\pi]$ (filled circles).
Empty circles display the ratio $\lambda_L^2 / \lambda_L^1$.
Statistical error in the $\xi_L(z)$ is $\simeq 5\%$. }
\end{figure}

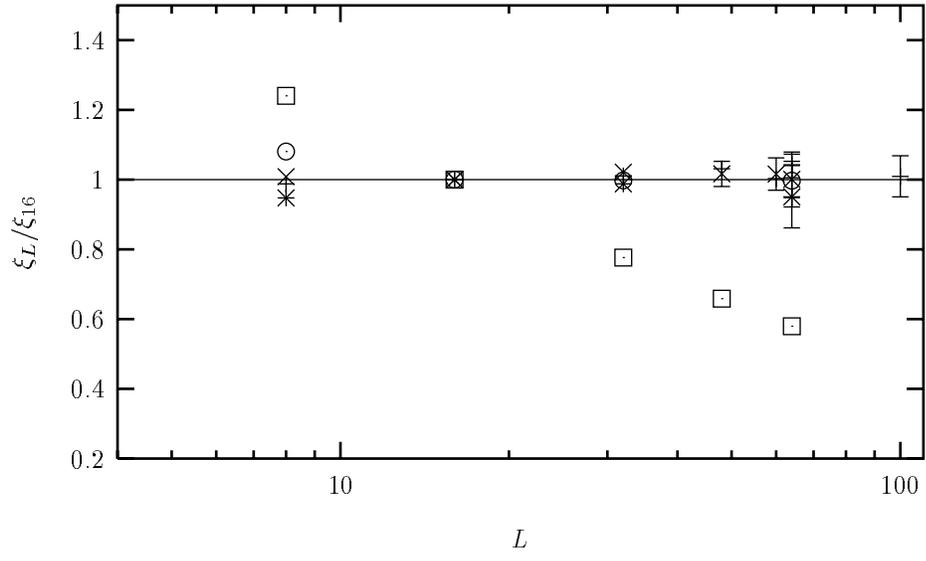
\begin{figure}
\caption{Scaled localization length, $\xi_L(z)/\xi_{16}(z)$,
plotted as functions of $L$ for various $z$.  +, X,
square, * and circle represent respectively $z=-1.0,-1.0+0.25i,
-1.0+0.6i, -1.25, -1.25+0.25i$. Error bars are displayed when
they exceed the symbol width. }
\end{figure}

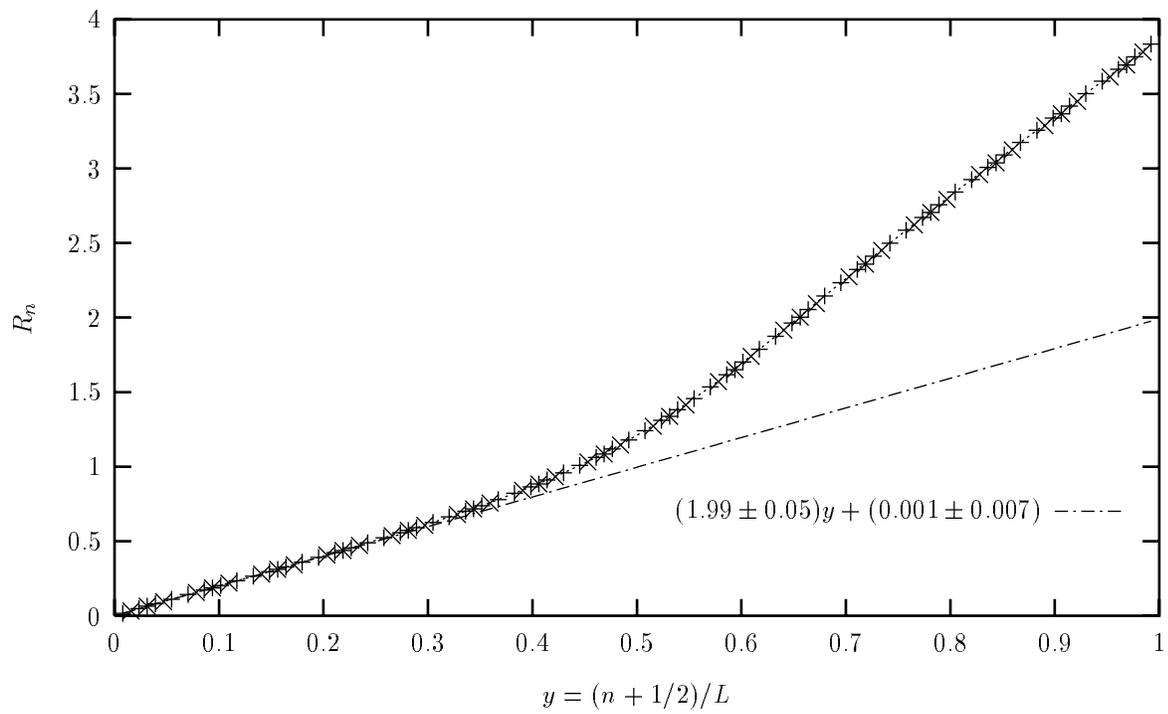
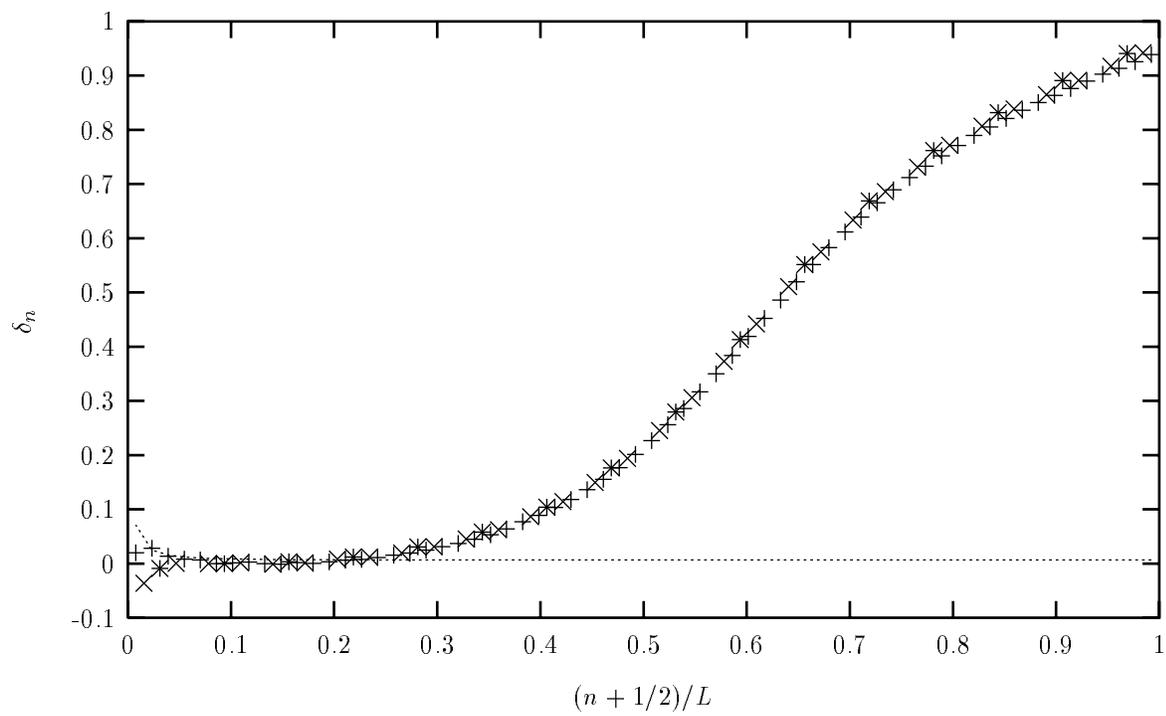
\begin{figure}
\caption{Comparison of the ratios, $R_n = \lambda_n/\lambda_{n_0}$,
for various different $L$ with equation (5) (dashed line).  The
largest correlation length corresponds to $n=0$.  A linear fit for
small $n/L$ yields $x=0.50\pm 0.02$.  To minimize the error in $R_n$
we choose a small $n_0$ for which the error is less than 1\%.
We collapse the ratios for different $L$ by defining $y=(n+1/2)/L$
and rescaling the ratio by $(2n_0 +1)/L$, where $n_0 \propto L$.
Parameters are:
$z=-1.0+0.25i$, $D={1 \over 4}$ and $n_0=8,4,2$ for $L=64$
(+), 32 (X) and 16 (*), respectively.  For $z=-1.0$
a similar picture is obtained with $x=0.49\pm 0.02$.  The inset
shows, for the same data set, the relative difference between the
numerically obtained values and equation (5).  The dashed line displays
the statistical error.}
\end{figure}

\end{document}